\documentclass[12pt,onecolumn]{IEEEtran}

\ifCLASSINFOpdf
  \usepackage[pdftex]{graphicx}
\else
\fi
\usepackage[cmex10]{amsmath}
\usepackage{amssymb}

\usepackage{epstopdf}

\usepackage{setspace}

\begin{document}

%
\title{\LARGE Comments on ``Capacity with explicit delay guarantees for generic sources over correlated Rayleigh channel"}

\doublespacing

%
%
%

\author{Ozgur~Ercetin,
        and~Mehmet~Ozerk~Memis,~\IEEEmembership{Student~Member,~IEEE}
\thanks{O. Ercetin and M. O. Memis are with the Faculty of Engineering and Natural Sciences, Sabanci University, Istanbul, 34956 Turkey e-mail: \{oercetin,ozerkmemis\}@sabanciuniv.edu.}
}

%
%

{}
%



\maketitle

\begin{abstract}
We address a major flaw in the abovementioned paper, which proposes to calculate effective capacity of random channels by the use of central limit theorem.  We analytically show that the authors are incorrect in finding the effective capacity by first taking the limit of cumulative random process rather than taking the limit of moment generating function of the same process. We later quantify our results over a correlated ON-OFF process.
\end{abstract}

\begin{IEEEkeywords}
Central Limit Theorem, moment generating functions, effective capacity.
\end{IEEEkeywords}

%
\IEEEpeerreviewmaketitle

\section{Introduction}

In \cite{wu-negi}, the authors inspired by the effective bandwidth theory have developed a dual effective capacity theory to analyze random and time-varying wireless channel under a probabilistic delay constraint.  The so-called \emph{effective capacity} provides a way to figure out the maximal constant arrival rate that can be sustained by the stationary ergodic service process, at the target Quality of Service (QoS) exponent.  The authors in \cite{beatriz-soret} claim that the method given in \cite{wu-negi} cannot be applied to general channel models such as time-correlated Rayleigh fading, and it is only parametrized with respect to the so-called QoS exponent.  Hence, they provide an alternate method of calculating the effective capacity which only involves the mean and variance of the cumulative service process.

We analytically show that the effective capacity calculated by the method in \cite{beatriz-soret} is wrong.  Furthermore, we numerically demonstrate that even when the proposed method is used as an approximation to the actual effective capacity, the approximation is only valid under certain channel conditions and for small values of QoS exponent.

\section{Brief Summary of Effective Capacity Theory}
For stationary ergodic arrival and service process, the queue-overflow probability is shown to be asymptotically decaying with exponential rate
\begin{align}
\lim_{x\to\infty}\log\left(\mathbb{P}\{Q(\infty)\}>x\right)=e^{-\theta x},
\end{align}
where $Q(\infty)$ is the queue length at stationary state, and $\theta$ is called the {\em QoS exponent}. The smaller $\theta$ is, the looser QoS guarantee achieves, which is reflected in the slower decay rate. On the contrary, faster decay rate with larger $\theta$ guarantees stringent QoS performance.

Let $c(\tau)$ be the instantaneous service rate of the queue in terms of bits that can be served in a finite length slot $\tau$, and $C(t)=\sum_{\tau=0}^{t}{c(\tau)}$ be the maximum aggregate number of bits that can be served during $[0,t]$.

Wu and Negi developed the concept of effective capacity \cite{wu-negi}, which is defined as the function of Gartner-Ellis (GE) limit of service process and the QoS exponent $\theta > 0$, i.e.,
\begin{align}
E_C(\theta)=-\frac{\alpha_C(-\theta)}{\theta}, \label{eq:eff-cap}
\end{align}
where $\alpha_C(\theta)$ is the GE limit of the service process.  GE limit of service process is defined in terms of the logarithm of moment generating function of cumulative service process $C(t)$.
\begin{align}
\alpha_C(\theta)=\lim_{t\to\infty}\frac{1}{t}\log\mathbb{E}\left[e^{\theta C(t)}\right]. \label{eq:alpha}
\end{align}

\section{Comments on the Analysis in \cite{beatriz-soret}}

In Section III of \cite{beatriz-soret}, the authors claim  that regardless of the distribution of instantaneous service process, $c(\tau)$, the effective capacity, $E_C(\theta)$,  is that of a Gaussian random variable with mean and variance depending on the service process.  We analytically show that in general this is not true, since the authors made a mistake while taking the GE limit of the cumulative service process.

For ease of exposition, we follow \cite{beatriz-soret}, and develop our arguments for the case of uncorrelated wireless channel, where $c(\tau), \forall \tau$, are independent and identically distributed (iid) random variables.  For this case, GE limit of service process can be found from \eqref{eq:alpha} as follows:
  \begin{align}
    \alpha_C(\theta) &=  \lim_{t \to \infty} \frac{1}{t} \log \mathbb{E}\left[e^{\theta C(t)}\right],\nonumber\\
    &=  \lim_{t \to \infty} \frac{1}{t} \log \mathbb{E}\left[e^{ \theta \sum_{\tau=0}^{t} c(\tau) }\right],  \nonumber \\
    &=  \lim_{t \to \infty} \frac{1}{t} \log \mathbb{E}\left[ \prod_{\tau=0}^{t} e^{\theta c(\tau)} \right], \nonumber \\
    &=  \lim_{t \to \infty} \frac{1}{t} \log \left(\prod_{\tau=0}^{t} \mathbb{E}\left[e^{\theta c(\tau)} \right] \right), \label{eq:alpha-uncorrelated}\\
    &= \lim_{t \to \infty} \frac{1}{t} \sum_{\tau=0}^{t} \log \mathbb{E}\left[e^{\theta c(\tau)} \right], \label{eq:alpha-correct}
  \end{align}
where \eqref{eq:alpha-uncorrelated} is obtained from the assumption that $c(\cdot)$ is iid.  Note that \eqref{eq:alpha-correct} shows that GE limit of cumulative service process is simply the limit of the sum of logarithm of moment generating function of the instantaneous service process $c(\cdot)$.

On the contrary, Soret et.~al., states in Section III of \cite{beatriz-soret} that as $t\to\infty$ central limit theorem can be applied, and $C(t)$ can be considered as a Gaussian random variable with average $t\cdot m_c$ and variance $t\cdot \sigma_c^2$, where $m_c=\mathbb{E}[c(\tau)]$, and $\sigma_c^2=\mathbb{E}\left[(c(\tau)-m_c)^2\right]$.  Note that the moment generating function of this Gaussian random variable is given as $M_C(\theta,t)=e^{\theta tm_c+\frac{\theta^2}{2}t\sigma_c^2}$. Even though, this argument is correct on its own, the derivation of effective capacity based on this argument appears flawed. Soret et.~al., give the logarithm of moment generating function of cumulative service process which is found to be a Gaussian random variable as:
\begin{align}
\tilde{\alpha}_C(\theta) &=\lim_{t \to \infty} \frac{1}{t} \log M_C(\theta,t)\nonumber\\
&=\theta\left(m_c+\frac{\theta}{2}\sigma_c^2\right).
\label{eq:alpha-wrong}
\end{align}
Clearly, \eqref{eq:alpha-wrong} and \eqref{eq:alpha-correct} are not equal to each other for all instantaneous service processes $c(\cdot)$, even when $t\to\infty$.  We demonstrate this with a simple example in the following section.

The problem with the argument made by Soret et.~al., is that the authors first take the limit of $C(t)$ which is not only inside both the logarithm and expectation but also appears in the exponent of Euler's number.  In addition, after moving the limit to the exponent of Euler's number, they still take the limit of moment generating function of Gaussian random variable divided by $t$.  In summary, \eqref{eq:alpha-wrong} is the result of the following mathematical statement:
  \begin{align}
    \tilde{\alpha}_C(\theta) &=\lim_{t_2 \to \infty} \frac{1}{t_2} \log \mathbb{E}\left[e^{{\theta \lim_{t_1 \to \infty} C(t_1)}}\right] \nonumber\\
    &=  \lim_{t_2 \to \infty} \frac{1}{t_2} \log \mathbb{E}\left[e^{\theta \lim_{t_1 \to \infty} \sum_{\tau=0}^{t_1} c(\tau)}\right]
  \end{align}
In general, $\alpha_C(\theta)$ and $\tilde{\alpha}_C(\theta)$ are not equal to each other for all $\theta$, and $c(\cdot)$.  In fact, a straightforward observation shows that these two quantities are equal to each other only when instantaneous service process $c(\cdot)$ is iid Gaussian random variable with mean $m_c$ and variance $\sigma_c^2$.

\section{A Numerical Example: On-Off Process}
One may argue that even if $\alpha_C(\theta)$ and $\tilde{\alpha}_C(\theta)$ are not exactly equal to each other, $\tilde{\alpha}_C(\theta)$ represents a good approximation to $\alpha_C(\theta)$ whose derivation is quite complex for a large variety of wireless channels.  In the following, we demonstrate that this is in fact not true, and the quality of approximation relies on the channel parameters as well.
For this purpose, we consider a {\em time-correlated} and slotted channel model, namely ON-OFF channel.    The main reason we consider ON-OFF channel model is that it is analytically sufficiently simple  so that exact closed form solution of $\alpha_C(\theta)$ can be obtained.  Meanwhile, ON-OFF channel still displays time-correlation among channel states similar to more complicated channel models such as Rayleigh channels. Note that the authors in \cite{beatriz-soret} argue that their approach is applicable not only to Rayleigh channel model but also to all other possibly correlated channel models.  By assuming ON-OFF channel model, we can explicitly and inarguably show that their method provides a good approximation to exact effective capacity under a limited range of QoS exponent $\theta$, and channel rates.

ON-OFF channel is modeled as follows.  In ON state, the user can send $r$ bits/slot, and in OFF state the user is not allowed to transmit any bits.  The transition probability from OFF(ON)-state to ON(OFF)-state is denoted by $1-\lambda$ ($1-\mu$), where $0\leq\lambda,\mu\leq 1$.  It is easy to determine that the stationary probability of being in ON-state is given as $\pi_{ON}=\frac{1-\lambda}{2-\lambda-\mu}$. GE limit of markov modulated process is found in \cite{cheng}.  Let $Q$ denote the transition probability matrix for an irreducible and aperiodic general $N$-state markov modulated process, and let $r_i$ be the service rate of each state $i, 1\leq i\leq N$.  Then,
\begin{align}
\alpha_C(\theta)=\log\left(\rho(Qe^{\theta R})\right),\label{alpha-MMP}
\end{align}
where $R=\mbox{diag}(r_1,r_2,\ldots, r_N)$ is a diagonal matrix of service rates and $\rho(A)$ is the spectral radius of matrix $A$. For ON-OFF source \eqref{alpha-MMP} simplifies to
\begin{align}
    \alpha_C(\theta) &= \log ( \frac{1}{2} a(\theta) + \frac{1}{2} \sqrt{a(\theta)^2 + 4 b(\theta)} ),
    \label{eq:alpha-ONOFF}
\end{align}
where
\begin{align}
    a(\theta) &= \lambda + \mu e^{r\theta} ,\\
    b(\theta) &= (1-\lambda-\mu) e^{r\theta}.
\end{align}
Note that effective capacity of ON-OFF channel can be determined by inserting \eqref{eq:alpha-ONOFF} into \eqref{eq:eff-cap}.

Meanwhile, in order to calculate $\tilde{\alpha}_C(\theta)$, we first need to calculate the mean and the variance of cumulative service process.  Soret et.~al., states that over a block of length $k$\footnote{$k$ is chosen to be sufficiently long to ensure that elements of different blocks can be considered approximately independent.}, the mean and variance of cumulative service process is given as
\begin{align}
\mathbb{E}\left[C(k)\right]&=km_c,\nonumber\\
\mbox{var}\left(C(k)\right) &=k\sigma_c^2+2\sum_{p=0}^{k-2}\sum_{q=p+1}^{k-1} K_c(q-p),\nonumber
\end{align}
where
\begin{align}
K_c(m)=\mathbb{E}\left[c(n)c(n+m)\right]-m_c^2.
\end{align}
For ON-OFF process described earlier,
\begin{align}
m_c=&r\pi_{ON}\nonumber,\\
\sigma_c^2=&r^2\frac{(1-\lambda)(1-\mu)}{(2-\lambda-\mu)^2}\nonumber\\
\mathbb{E}\left[c(n)c(n+m)\right]=&r^2\pi_{ON}\frac{1-\mu+(1-\lambda)(\lambda+\mu-1)^m}{2-\lambda-\mu}.\nonumber
\end{align}
According to the method given in \cite{beatriz-soret}, $\tilde{\alpha}_C(\theta)$ can be determined as
\begin{align}
\tilde{\alpha}_C(\theta)=\theta\left(m_c+\frac{\theta}{2}\frac{\mbox{var}\left(C(k)\right)}{k}\right)
\end{align}

\begin{figure*}[ht]
\begin{minipage}[b]{0.5\linewidth}
\centerline{\includegraphics[width=3in]{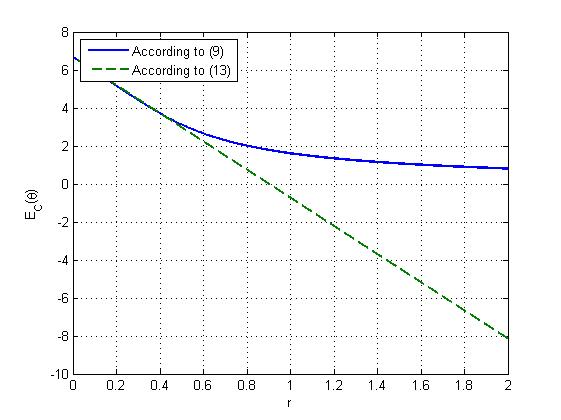}}
    \caption{Effective capacity of ON-OFF source with respect to QoS exponent $\theta$ when $r=10$, $\lambda=0.2$, and $\mu=0.6$.   \label{fig:theta}}
\end{minipage}
\hspace{0.5cm}
\begin{minipage}[b]{0.5\linewidth}
\centerline{\includegraphics[width=3in]{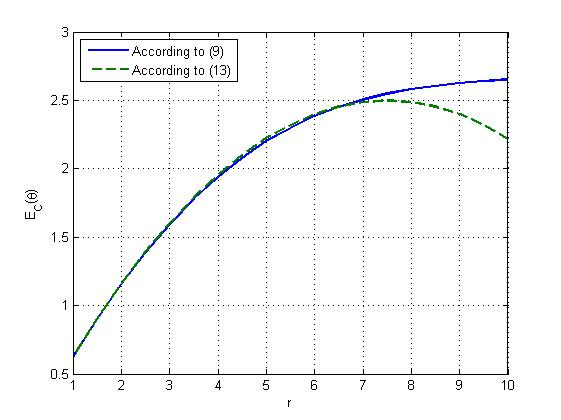}}
    \caption{Effective capacity of ON-OFF source with respect to transmission rate $r$  when $\theta=0.6$, $\lambda=0.2$, and $\mu=0.6$. \label{fig:r}  }
\end{minipage}

\end{figure*}

We compare effective capacity calculated with $\alpha_C(\theta)$ and $\tilde{\alpha}_C(\theta)$ with respect to $\theta$ and $r$.  The state transition probabilities of ON-OFF source are arbitrarily chosen as $\lambda=0.2$, and $\mu=0.6$.  Note that we achieved similar results for other values of $\lambda$, and $\mu$.  In Figure \ref{fig:theta}, we observe that effective capacity calculated with $\alpha_C(\theta)$ and $\tilde{\alpha}_C(\theta)$ match well when the range of $\theta$ is in $[0,0.4]$.  However, one can easily notice that these two lines deviate from each other as $\theta$ increases.  When $\theta$ is larger, i.e., QoS guarantees are stricter, then $\tilde{\alpha}_C(\theta)$ gives incorrect negative effective capacity values.

In Figure \ref{fig:r}, we compare the effective capacities with $\alpha_C(\theta)$ and $\tilde{\alpha}_C(\theta)$ for varying values of transmission rate $r$.  Again, we clearly observe that the approximation given in \cite{beatriz-soret} is only correct for small values of $r$.


\section{Conclusion}
In this work, we analytically show that in \cite{beatriz-soret} the authors make a major flaw in finding the effective capacity by first applying the central limit theorem to the sum of iid random variables $c(\tau)$ with mean $m_c$ and variance $\sigma_c^2$, and then taking the GE limit of the resulting cumulative random process. Next, over a correlated ON-OFF channel we numerically verify that $\tilde{\alpha}_C(\theta)$ is not a good approximation to $\alpha_C(\theta)$ either, since the quality of approximation depends to a great degree on the channel parameters. Besides, given the channel parameters the proposed effective capacity expression of \cite{beatriz-soret} tends to follow the exact solution only in a limited range for QoS exponent $\theta$ and negative capacity values are achievable as seen in Figure \ref{fig:theta}. The same observation is also true for the transmission rate $r$ as shown in Figure \ref{fig:r}.  Based on these observations, we caution the researchers on the applicability of the approach in \cite{beatriz-soret}, and recommend them to verify that the channel and QoS parameters are chosen so that the approximation is correct. \vfill

\break

\ifCLASSOPTIONcaptionsoff
  \newpage
\fi

\section*{Acknowledgment}
This work was supported in part by TUBITAK grant number 109E242.



%

\end{document}